# Thermodynamic investigations in the precursor region of FeGe

**L. Cevey[1], H. Wilhelm[2], M. Schmidt[3], and R. Lortz[1]**

[1]Department of Physics, The Hong Kong University of Science & Technology, Clear Water Bay, Kowloon, Hong Kong
[2]Diamond Light Source Ltd., Chilton, OX11 0DE, United Kingdom
[3]Max Planck Institute for Chemical Physics of Solids, Nöthnitzer-Str. 40, 01187 Dresden, Germany

High-resolution DC magnetization and AC-specific heat data of the cubic helimagnet FeGe have been measured as function of temperature and magnetic field. The magnetization data as well as the isothermal susceptibility data confirm the complexity of the magnetic phase diagram in the vicinity of the onset of long-rang magnetic order ($T_c$ = 278.5 K) and the existence of a segmented A-phase region. Moreover, these data revealed independent and clear indications of phase boundaries and crossovers within the A-phase region. Together with the anomalies in the specific-heat data around $T_c$ and at small magnetic fields ($H$ < 600 Oe) a complex magnetic phase diagram of FeGe is obtained.

## 1 Introduction

Cubic helimagnets crystallizing in the non-centrosymmetric B20 structure (space group $P2_13$) have been long known for their peculiar magnetic properties [1-3]. These compounds, mainly transition-metal silicides or germanides (TM = Mn, Fe, and Co), display chiral helical and skyrmionic modulations with periods up to several hundred unit-cells [2-4] which is well understood [5-8]. In a narrow region in the magnetic field ($H$) – temperature ($T$) phase diagram, however, the subtle balance of isotropic and much weaker anisotropic interactions such as exchange anisotropy and cubic magneto-crystalline anisotropy lead to yet not well understood anomalies in the vicinity of the onset of long-range magnetic order [9,10]. This *precursor region* comprises the so-called A phase and a sequence of complex phase transitions and crossovers in the vicinity of the onset of long-range magnetic order at $T_c$. These precursor phenomena are related to the softening of the magnetization modulus and a strong coupling between angular and longitudinal modes near $T_c$ [10-12]. It has been shown [8,10] that the crossover of inter-core interactions and the onset of specific confined helical and skyrmionic states lead to a complex phase diagram with different chiral modulations [9].

Here we present high-resolution thermodynamic measurements on single crystalline FeGe which are complementary to previously reported magnetic ac-susceptibility data [9]. The complexity of the magnetic phase diagram around $T_c$ and the segmentation of the A-phase region are confirmed. Therefore, the idea of interpreting the A phase as a certain distinct single-modulated phase with either one-dimensional [13,14] or 'triple-q' modulations [15-17] has to be rejected.

## 2 Experimental Details

The specific heat, $C_p$, was measured in a home-made AC Peltier modulated-bath calorimeter [18]. It allows us to measure $C_p$ during both, temperature and field scans. The magnetization was measured in a Quantum Design Vibrating-Sample SQUID magnetometer (VSM) in continuous sweeping mode. Magnetic susceptibility was obtained by taking the field derivative of the magnetization data. The orientation of the sample to the field direction was arbitrary but identical in both experiments.

## 3 Results

Figure 1 shows the field dependence of the isothermal susceptibility $\chi(H)$ of FeGe for temperatures around $T_c$ = 278.5 K. The initial increase in low fields is associated with the transition from a helical to the cone structure at $H_{c1}$ [9]. In the conical phase the susceptibility is field independent and shows a plateau for $T$ < 274 K. For temperatures 274.5 < $T$ < 278.5 K, however, a minimum in $\chi(H)$ evolves. This is the fingerprint of the 'A-phase region' (see a discussion and bibliography in [10]). Further increase of the field leads to the spin-flip transition into the field-polarized (FP) state at the upper critical field $H_{c2}$. These are the generic phase transitions observed in FeGe [9,10] and other B20 magnets [19-22]. However, the present high-resolution measurements provide further details in the precursor region.

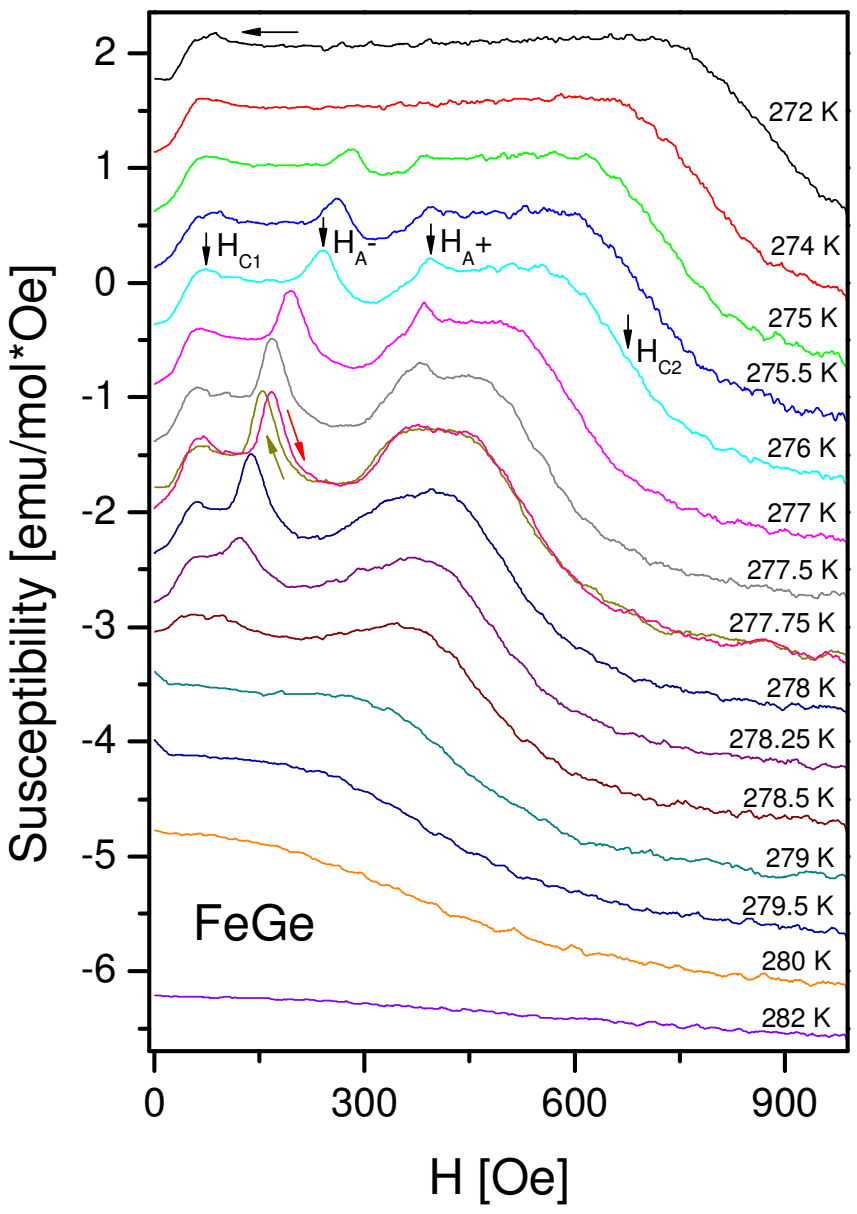

**Figure 1** Isothermal $\chi(H)$ data of FeGe with arbitrary field direction. The curves are stacked with a -0.5 emu/mol/Oe offset. Data at 277.75 K were recorded upon increasing and decreasing fields. The small hysteresis indicates that the transitions are of first-order.

The spin-flip transition into the FP state is not affected by the appearance of the A-phase region, thus underlining its pocket nature within the cone structure. The peak present in $\chi(H)$ for 275 K $\leq T \leq$ 277.5 K and delimitating the high-field side of the valley at a characteristic field $H_A^+$ disappeared at 277.75 K. Instead, it is followed by a gradual reduction of $\chi(H)$ and a small plateau at higher fields ($T$ = 277.75 K). This indicates that a cone-like-structure is stabilized before entering the FP state. A different magnetic state, however, seems to be present at 278 K as the plateau in $\chi(H)$ has transformed into a bump which persists up to $T_c$.

The large peak on the low-field side of the valley at $H_A^-$ is present above 274 K and smears out at $T_c$. Thus, only a transition from the cone to the A-phase region occurs. The presence of three sharp peaks supports a first-order nature of the transitions between the helical, conical and the A phase. This is also supported by the small hysteresis of about 10 Oe in the transition field observed at 277.75 K for both field-sweep directions.

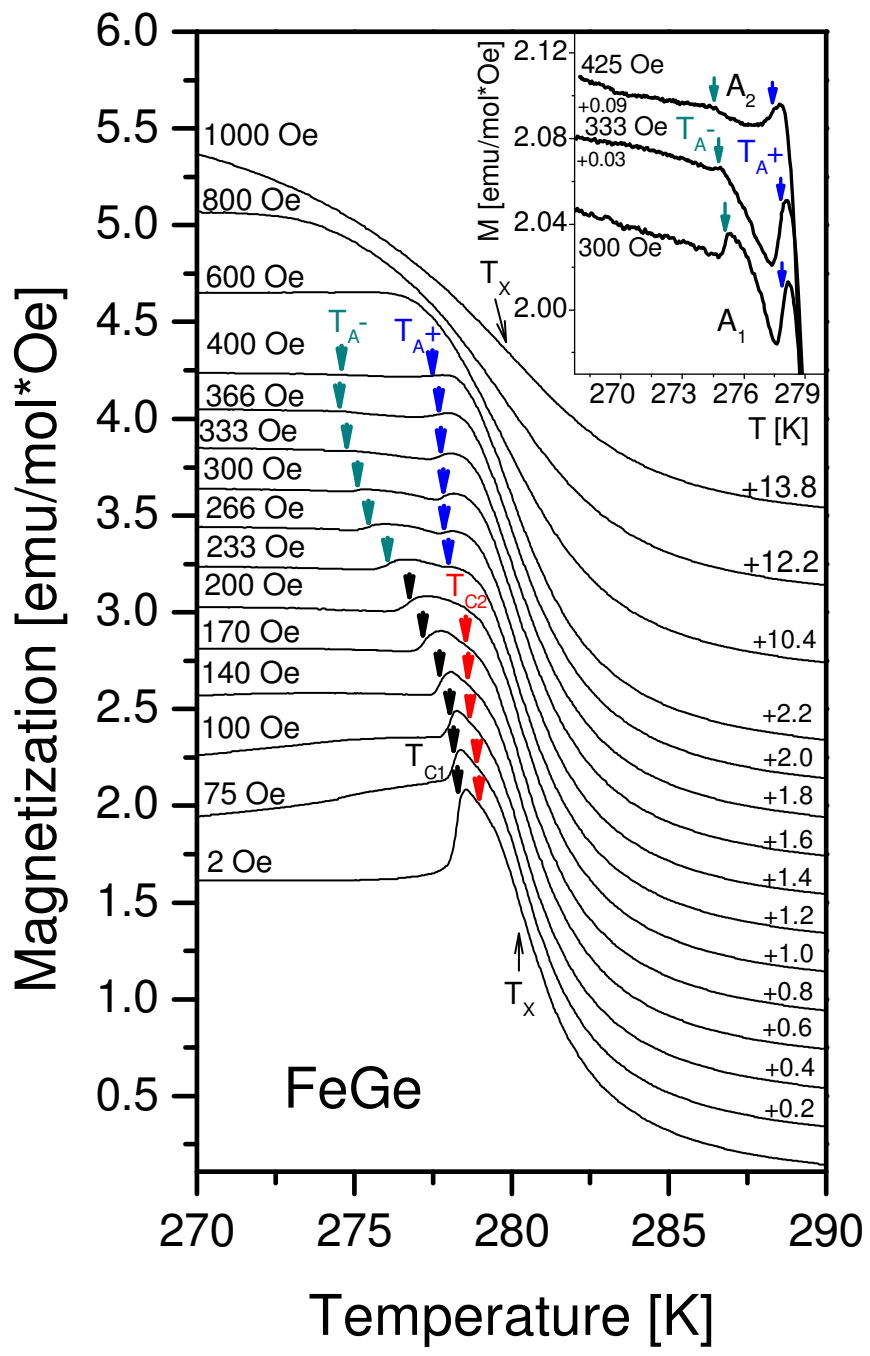

**Figure 2** $M(T)$ data normalized by the applied field of FeGe at fixed applied fields $0 < H < 1000$ Oe. All curves are shifted with respect to the 2 Oe data as indicated. The inflection points marked by arrows indicate the various magnetic transitions. Inset: Details of the magnetization data showing distinct anomalies at the transitions into the $A_1$ phase ($H$ = 300 and 333 Oe) and the $A_2$ phase (425 Oe). The curves for $H$ = 333 and 425 Oe are shifted relative to the 300 Oe data.

Zero-field-cooled temperature scans of the magnetization, $M(T)$, normalized by the corresponding field, are shown in Figure 2. Comparison with field-cooled data did not show significant differences. The main features of the susceptibility described above are all observed in the $M(T)$ data. The overall increase in magnetization for decreasing temperatures is associated with the appearance of some kind of magnetic ordering at a crossover temperature $T_x$, defined as the inflection point in $M(T)$. $T_x$ hardly changes and is located at about 280 K. At lower temperature a pronounced cusp marks the transition into the helical state ($H$ = 2 Oe), with the magnetization stabilizing at a constant value below the maximum. Its position at $T_{c1}$ = 278.5 K is defined as the inflection point of the $M(T)$ data. The cusp is preserved upon increasing field, though the drop in magnetization largely decreases upon entering the cone structure for $H$ > 100 Oe. It gradually becomes an individual anomaly while shifting to lower temperatures. A second maximum can be identified for fields around 250 Oe. The inflection points at these two features are taken as the lower and upper end of the A-phase region $T_A^-$ and $T_A^+$ (blue and dark cyan arrows), respectively. A kink right above the peak at low fields may indicate the transition into some intermediate phase between the helical and the paramagnetic (PM) phase, and is thus marked as $T_{c2}$ (red arrows). From the continuous evolution of the main cusp, it seems that the A phase somehow joins the helical phase around $T_{c1}$.

A segmentation of the A-phase region into two parts becomes obvious from the $M(T)$ data above 300 Oe (see inset to Fig. 2). The minimum in $M(T)$ and the cusps associated with the $A_1$ phase becomes smaller with increasing field and transforms into a shallow minimum between 400 Oe and 500 Oe without any pronounced kink-like anomalies at the low- and high-temperature side. This is taken as indication of a different magnetic phase ($A_2$) within the A-phase region, as observed previously [9]. Based on the magnetization data it can be inferred that its magnetic characteristics differs only little from the conical phase but seems to be definitely different from the $A_1$ phase.

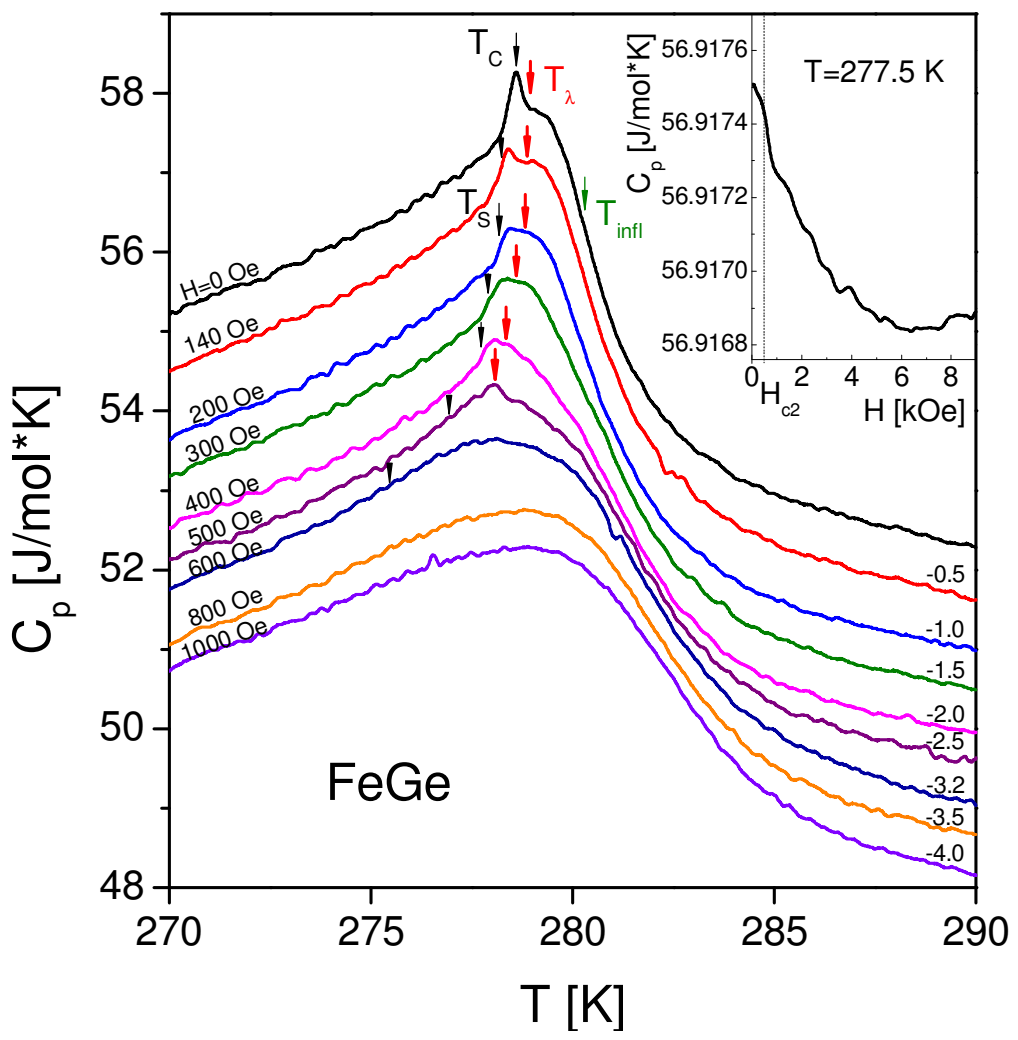

**Figure 3** AC specific-heat $C_p(T)$ of FeGe in various fields. All curves are shifted relative to the zero-field data. The meaning of the various arrows is described in the text. Inset: Specific heat during a field sweep at $T$=277.5 K. A strong drop is observed at the upper critical field $H_{c2}$.

Figure 3 shows the high-resolution ac-specific-heat data which reveal further details of the evolution of the various magnetic transitions and cross-overs. In zero field, a sharp peak is visible at $T_c$ = 278.5 K, superimposed slightly below the top of a λ-like anomaly. This extraordinarily sharp anomaly highlights the excellent quality of the sample and sample inhomogeneity can be excluded. The peak at $T_c$ corresponds to the transition into the helical phase. Its size is reduced at 140 Oe while it slightly shifted to 278.4 K, and finally evolved into a step at temperature $T_S$ and fields above 200 Oe. Its inflection point, present at least until 600 Oe, is defined as $T_S$. The evolution of the sharp peak at $T_c$ into a step at $T_S$ corresponds to the transition into the A-phase region. On the other hand, the main λ-like anomaly becomes weaker, broadens and changes its shape for $H$ > 140 Oe. The inflection point at $T_{infl}$ in $C_p(T)$ above $T_\lambda$ is taken as criterion for the center of the crossover into the PM state.

Above $H$ = 600 Oe only a very broad peak is observed. Its position moves towards higher temperature with increasing field. This is the typical behavior for the high-field crossover region of a magnetic transition from the PM phase to the FP state. Isothermal field-sweep specific heat data, $C_p(H)$ at 277.5 K across the A-phase region is shown in the inset to Figure 3. The relative signal variation was found to be a tiny fraction of the large total specific-heat value ($\Delta C_p$ = 25 $10^{-6}$ J/mol·K). Nonetheless, the strongest decrease in $C_p(H)$ could be repeatedly resolved and correlates with $H_{c2}$. The overall drop in $C_p(H)$ reflects the transitions from the magnetically modulated phase region to the well-defined FP state at about 6 kOe.

## 4 Discussion

The various features found in the susceptibility, magnetization and specific heat data are plotted in the $(H,T)$ phase diagram shown in Figure 4. Although the DC magnetization reveals more pronounced anomalies at the boundaries of the $A_1$ phase, the derived phase diagram is very similar to that reported in [9,10].

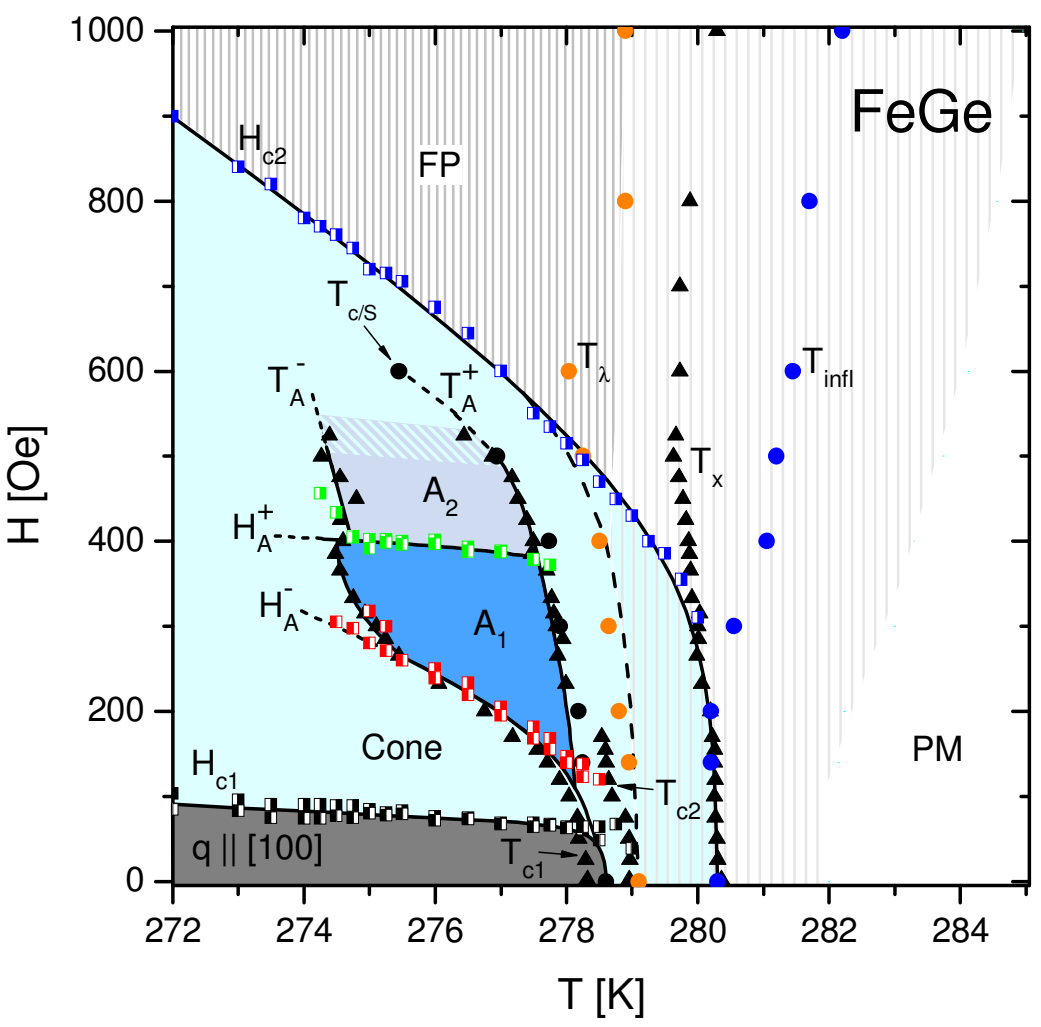

**Figure 4** Phase diagram of FeGe in the vicinity of magnetic ordering ($T_c$ = 278.5 K) obtained from the susceptibility, magnetization and specific-heat data presented in section 3. The various transition lines and crossover temperatures are displayed with the following symbol codes: half-full squares for $\chi(H)$ (filling on the right half and left half mark data taken for increasing and decreasing field, respectively), triangles for $M(T)$ and circles for $C_p(T)$. The helical ground state transforms into a conical phase at a lower critical field $H_{c1}$. The field-polarized (FP) state is entered at the upper critical field $H_{c2}$ for $T < T_c$. A segmented A-phase region with pockets $A_1$ and $A_2$ is embedded in the conical phase. The grey shading marks qualitatively the region over which the broad crossover extends with the high-temperature limit estimated from the onset of the specific-heat anomaly.

The specific-heat data are in excellent agreement with the magnetization, bringing a further confirmation for the helical phase and the upper transition line of the A phase: The observed tiny peak-like anomaly superimposed onto the λ-like transition marks the upper end of the cone phase and in higher fields when it transforms into a step it represents the upper end of the A-phase region. Meanwhile, the inflection point on the right side of the main λ-like transition is in excellent agreement with the $T_x$ line from the magnetization temperature scan. The shape of the magnetic specific-heat transition resembles a broadened λ-like anomaly, typical for a transition with strong fluctuations [10].

Our data clearly verifies the separation of the A phase in multiple pockets: the $A_1$ and the $A_2$ phases exhibit different magnetic properties and are bounded by two first-order and two second-order transitions, respectively. The sharp features in $M(T)$ related to the $A_1$ and $A_2$ transition are a consequence of a different magnetic structure and are not caused by micro-stress as suggested in [21]. The step-like cusps forming the $T_{c1}$ / $T_A^-$ and $T_A^+$ lines again seem to support the discontinuous nature of the transition into the helical and $A_1$ phases, while the kink-like anomalies at the transition into the $A_2$ phase suggests a continuous second-order nature. It remains unclear what happens exactly between the A region's upper transition $T_A^+$ and the $T_x$ line. Although the magnetization data $M(T,H=0)$ indicate a completion of magnetic ordering right upon entering the helical state, the specific heat shows a large anomaly centered at temperatures slightly above $T_c$, and entering the helical phase appears as a sharp additional superimposed peak.

Comparing the previous AC susceptibility data [9] with the present data illustrates that AC susceptibility underestimated the size of the anomalies of the first-order transitions. The reason is certainly the associated meta-stability. The magnetization data presented here resemble remarkably to data presented for MnSi [22]. Indeed these data are consistent with earlier findings pointing to a complex A-phase region in MnSi as well [19]. The segmentation of the A-phase region is a generic feature and not restricted to binary transition-metal silicides or germanides as the recent data on $Cu_2OSeO_3$ show. The observation of two ESR lines in the center of the A phase [20] and a change of the propagation direction therein, found by small angle neutron scattering [21], are clear indications that this phase space in $Cu_2OSeO_3$ is inhomogeneous as well.

## 5 Conclusions

A narrow temperature-magnetic field region around $T_c$ = 278.5 K of the cubic helimagnet FeGe was investigated with high-resolution DC magnetization and AC specific-heat measurements. The deduced phase boundaries and crossover lines are in perfect agreement with those obtained from AC susceptibility data [9] and confirm the separation of the A phase into several pockets. Thus, these data unambiguously confirm the existence of a multiplicity of magnetic phases within the precursor region in FeGe which is characterized by complex magnetic modulations and can neither be explained by a distinct single-modulated nor a simple triple-helix magnetic structure.

**Acknowledgements** This work was supported by the Research Grants Council of Hong Kong Grant SEG_HKUST03.